# CMOS-compatible, multiplexed source of heralded photon pairs: towards integrated quantum combs

Christian Reimer,[1] Lucia Caspani,[1,*] Matteo Clerici,[1,2] Marcello Ferrera,[1,2] Michael Kues,[1] Marco Peccianti,[1,3] Alessia Pasquazi,[1,3] Luca Razzari,[1] Brent E. Little,[4] Sai T. Chu,[5] David J. Moss,[1,6] and Roberto Morandotti[1]

[1]*INRS-EMT, 1650 Boulevard Lionel-Boulet, Varennes, Québec J3X 1S2, Canada*
[2]*School of Engineering and Physical Sciences, Heriot-Watt University, SUPA, Edinburgh EH14 4AS, UK*
[3]*Department of Physics and Astronomy, University of Sussex, Falmer, Brighton BN1 9QH, UK*
[4]*Xi'an Institute of Optics and Precision Mechanics of CAS, Xi'an, China*
[5]*City University of Hong Kong, Department of Physics and Material Science, Tat Chee Avenue, Hong Kong, China*
[6]*School of Electrical and Computer Engineering, RMIT University, Melbourne, VIC, 3001 Australia*
*\*caspani@emt.inrs.ca*

**Abstract:** We report an integrated photon pair source based on a CMOS-compatible microring resonator that generates multiple, simultaneous, and independent photon pairs at different wavelengths in a frequency comb compatible with fiber communication wavelength division multiplexing channels (200 GHz channel separation) and with a linewidth that is compatible with quantum memories (110 MHz). It operates in a self-locked pump configuration, avoiding the need for active stabilization, making it extremely robust even at very low power levels.

## 1. Introduction

With the realization of commercial solutions for quantum cryptography, research on integrated sources of entangled and single photons has been a hot topic in photonics in recent years. On the other hand, quantum logic ports and circuits for quantum computing algorithms have been demonstrated in integrated photonic chips [1,2]. It is foreseeable that on-chip sources of single and entangled photons at telecom wavelengths will represent a key enabling technology for quantum communication and computing [3], thus motivating an intense research activity [4–10]. Among the different approaches proposed so far, microcavities [11] are a very promising solution because of the field enhancement that they provide (for example the pair production rate for a $\chi^{(3)}$ triply resonant cavity scales as the sixth power of the cavity enhancement factor [12]).

Furthermore, practical long-distance quantum key distribution (QKD) [13], as well as quantum networks in general [14], and other applications such as information storage during quantum computation [15], will require quantum repeaters and memories. These devices are typically based on atomic transitions that have linewidths on the order of 10 to 100 MHz [13]. Exploiting optical cavities for the generation of photon pairs allows to reduce the photon pair bandwidth [16,17], eventually matching the quantum memories requirements without the need of narrowband filtering, which typically results in a reduced pair production rate. However, many of the sources based on integrated resonators have so far failed to achieve the narrow linewidths compatible with quantum memories because of their relatively modest Q-factors [6,8,18,19]. While narrow linewidth sources have been realized using extremely high Q-factor cavities, these are fundamentally incompatible with large-scale integration [5,7,20–22].

In addition, practical sources for QKD applications should be compatible with commercial optical communications networks. To this purpose, the possibility of generating many independent photon pairs simultaneously at different wavelengths matching the standard International Telecommunication Union (ITU) fiber communication channels will allow the realization of frequency multiplexed approaches in order to accommodate multiple users over the same fiber. This, in addition, is a necessary constraint for exploiting novel high-dimensional QKD protocols [23,24]. Multiplexed quantum photon pair generation is thus one of the grand technical challenges in quantum optics, which has only just recently started to be addressed [19,20,22]. Finally, and very importantly, microcavities (for both classical and quantum applications) are almost always pumped with an external laser that requires active stabilization such as thermal locking in order to achieve even short metastable operation [25,26]. However, as we have experimentally observed, this approach becomes ineffective at the low powers typically employed for photon pair generation, especially when microcavities with the high Q-factors required by quantum memories are considered.

In this paper, we proposed and characterized an integrated photon pair source that generates multiple, simultaneous, and independent photon pairs multiplexed and distributed on a frequency comb grid that is compatible with the ITU channel spacing of optical fiber communications networks. It also features linewidths that are orders of magnitude narrower than previous sources based on integrated ring resonators, and that are thus compatible with atomic-based quantum memories. Furthermore, it employs a self-locked pumping scheme that *does not* require an external pump laser, and hence any active stabilization. This makes it intrinsically stable even at the very low (sub-threshold) pump power levels typically used for photon pair generation. Finally, the proposed device is based on a microring resonator fabricated on a platform that is compatible with electronic computer chip technology (CMOS – for a detailed discussion see [27]). All together these features potentially mark a substantial step forward to achieve stable, integrated and CMOS-compatible photon pair sources for quantum optical applications.

## 2. Experiment

*2.1 Self-locked pumping scheme and experimental setup*

Figure 1(a) shows the experimental configuration. The ring resonator is a four-port device with a Q-factor of 1.375 million (17.9 field enhancement factor), 140 MHz linewidth, and a free spectral range (FSR) of 200 GHz (135 μm radius). The microring is vertically coupled to bus waveguides (1.45 μm x 1.5 μm core) and is fabricated using UV photolithography and reactive ion etching in a CMOS-compatible high refractive index doped glass platform – Hydex [27], featured by very low linear and negligible nonlinear optical losses, and high effective nonlinearity. The dispersion is small and anomalous, guaranteeing a large four-wave mixing (FWM) gain bandwidth at telecom wavelengths. The input and output bus waveguides are pigtailed to standard single mode fibers, resulting in coupling losses of <1.5 dB per facet (more details can be found in [28,29]). The ring resonator has resonances that are aligned to the ITU grid on a 200 GHz spacing. We note that this is achievable without any need for post-fabrication tuning.

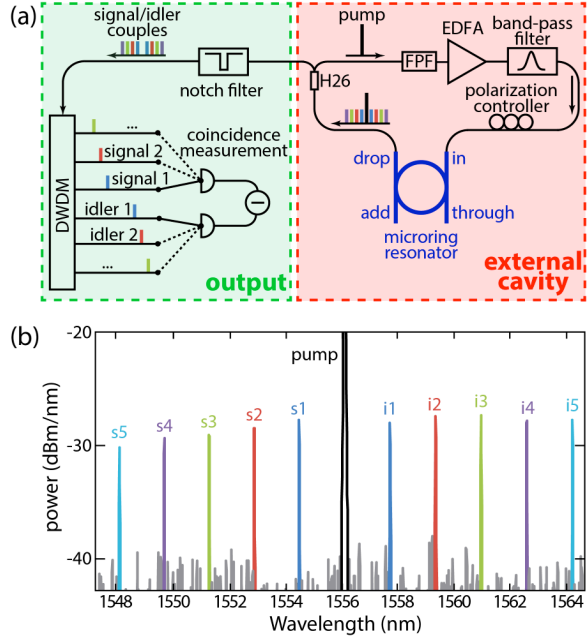

Fig. 1. (a) Experimental setup. The microring is embedded in an external cavity including a gain medium (EDFA), a band-pass filter (centered at the pump wavelength), and a polarization controller. Lasing is initiated by the ASE of the amplifier, which is fed back to the amplifier by means of 1x2 DWDM filter (H26 in the figure). The Fabry-Perot filter (FPF) selects only one of the external cavity lines. The signal/idler photon pairs exiting the drop port are transmitted through the high-isolation notch-filter, separated by a commercial DWDM filter, and then characterized by coincidence detection. (b) Spectrum of the above-threshold OPO showing the five different signal/idler (s5-s1 and i1-i5, respectively) couples detected in our experiment. The black curve represents the pump, while the grey traces are due to the noise of the optical spectrum analyzer.

In order to avoid the use of an external pump laser, we exploit a scheme we recently introduced in the context of a classical optical parametric oscillator (OPO) [30] that consists in embedding the microring in an external active cavity [Fig. 1(a)]. Lasing is initiated by the amplified spontaneous emission (ASE) from the erbium doped fiber amplifier (EDFA) that acts as the gain medium for the pump laser, which is then sustained by re-injecting one particular wavelength among the ring resonances (identified as the pump) into the amplifier. We note that the resonance shift experienced by microring resonators can be significant – on the order of hundreds of resonance linewidths [25]. For example, Hydex shows a thermal shift of –2 GHz/°C. However, the key feature of the proposed scheme is that the pump is automatically self-locked to the ring resonance without the need of any external tuning or feedback loop [30,31].

The band-pass filter selected a single ring resonance in order to achieve continuous wave (CW) pumping, which is preferable for generating high-coherence photon pairs. The output from the amplifier was filtered by a high-isolation (>180 dB) band-pass filter (Lightwave2020), centered at the pump wavelength, in order to remove the ASE at the signal/idler wavelengths. A fiber polarization controller was then used to adjust the pump polarization to the TE modes of the microring. When pumped below threshold, the OPO generated individual photon pairs on a frequency comb determined by the FSR of the ring resonator. The output of the OPO was then filtered by a 1x2 dense wavelength division multiplexing (DWDM) filter that reflected the pump into the amplifier, thus closing the external cavity loop in order to sustain lasing. The signal/idler pairs were transmitted to the detection stage, where the pump wavelength was further suppressed by means of a notch filter (90 dB isolation), while the frequency comb was de-multiplexed using a standard telecommunications DWDM filter (Optiworks) in order to separate the different wavelengths. Photon pairs at a particular wavelength were then detected by connecting the different outputs of the DWDM filter to single photon detectors. The total loss for the signal and idler arms are estimated to be 10.9 dB and 10.4 dB (excluding the quantum efficiency of the detectors), respectively. Of this, 6 dB is the intrinsic loss due to the fact that we collected the photon pairs only at one port (drop), 1.5 dB is due to the coupling from the chip to the fiber, 0.4 dB is due to the 1x2 DWDM reflecting the pump, 2 dB is due to the notch filter, and 0.6dB (1dB) is due to the DWDM filter in the idler (signal) arm. We then performed coincidence measurements exploiting two single photon detectors (idQuantique id210) set to free-running mode, 5% quantum efficiency, 25 μs dead time (to limit false counts due to after pulse effects). With these settings the dark count rates were 1.3 kHz and 3.4 kHz, respectively. The output of both detectors was then connected to a time-to-digital-converter (idQuantique id800, set to 81ps time-bins) for recording the coincidence histogram.

The EDFA used in our experiments (Amonics AEDFA-33-BFA, 30dB of gain at 0dBm, 30m active fiber) had a relatively long active length which resulted in a small FSR of the external pump cavity (6 MHz), and hence a large number of external lines (~24) lasing within a single microring resonance. This led to a chaotic pulsed-pump configuration [30] that, while interesting, is not ideal for high-dimensional QKD applications due to the resulting low coherence time. It has indeed to be considered that a long pump coherence time corresponds to a large alphabet, i.e. high number of bits of information per photon [32]. To address this issue we introduced a narrowband tunable fiber Fabry-Perot Filter (FPF) with 1 MHz bandwidth and 200 MHz FSR inside the laser cavity. We note that once the FPF was tuned to achieve highest pump powers and stable operation, no further stabilization was necessary. The FPF allowed the selection of a single external cavity line per ring resonance, thus suppressing the chaotic dynamics, and producing stable CW oscillation of the pump. We note that the issue of stability, i.e. the locking between the external pump laser and the ring resonances, is much more relevant for the high Q-factor microcavities needed to achieve high generation efficiencies and also to match the narrow linewidths required for atomic-based quantum memories. For very high Q-factor ring resonators, such as the one employed in this work, the locking between the external CW laser and the ring resonance is usually lost after a few minutes, especially at the low pump powers typically used for photon pair generation. In contrast, the considered self-locked configuration is extremely stable and reliable, allowing continuous operation of the quantum source over several days with pump power fluctuations lower than 5%. We note that no active stabilization was required in any of the components of the setup (microring, FPF, DWDM filters, etc.).

*2.2 Multiplexed photon pair generation*

We demonstrated the multiplexed nature of the proposed integrated photon pair source by pumping the ring well below the OPO threshold (30 mW at the ring input port, with the OPO threshold being around 120 mW), and selecting five different signal/idler couples symmetrically located around the pump wavelength (see Fig. 1(b) and the list in Table 1). Clear coincidence peaks are visible on all five symmetric channel pairs (Fig. 2(a)), while no coincidences are measured between non-diagonal elements of the frequency matrix (Fig. 2(b)).

For each signal-idler channel pair we evaluated the coincidence to accidental ratio (CAR) following the procedure described by Engin et al. [18]. We consider the sum of the coincidence counts (CC) within the full width at half maximum (FWHM) of the coincidence peaks, while we evaluate the accidental counts (AC) in a window of the same width far outside the peak. The CAR is then evaluated as the ratio CC/AC. For a pump power of 30 mW at the ring input we obtained CAR values between 10 and 14, single rates between 8.5 and 10 kHz per channel, and coincidence rates between 26.4 and 48.4 Hz per channel (raw values, without any background subtraction).

**Table 1. Signal and idler wavelengths corresponding to the DWDM channels of the ITU grid, commonly exploited in standard optical communications**

|  | DWDM Channel | Wavelength (nm) |
|---|---|---|
| Signal 5 | H36 | 1548.11 |
| Signal 4 | H34 | 1549.72 |
| Signal 3 | H32 | 1551.32 |
| Signal 2 | H30 | 1552.93 |
| Signal 1 | H28 | 1554.54 |
| **Pump** | **H26** | **1556.15** |
| Idler 1 | H24 | 1557.77 |
| Idler 2 | H22 | 1559.39 |
| Idler 3 | H20 | 1561.01 |
| Idler 4 | H18 | 1562.64 |
| Idler 5 | H16 | 1564.27 |

Considering the losses of the detection system (but not the on chip and coupling losses) and the dark counts of the detectors, this corresponds to a pair production rate per channel between 286 and 346 kHz (at the output port – drop – of the ring), and to a pair production probability per channel between 2.4 and 2.9 × $10^{-12}$ (measured as the probability that two pump photons at the input fiber of the ring produce a signal/idler pair at the output fiber of the ring). While this might appear low, we note that this output rate is evaluated for a single channel. The total pair production rate for all channels (see Sec. 3) can be estimated to be ~25 MHz (considering a nearly constant generation rate per channel).

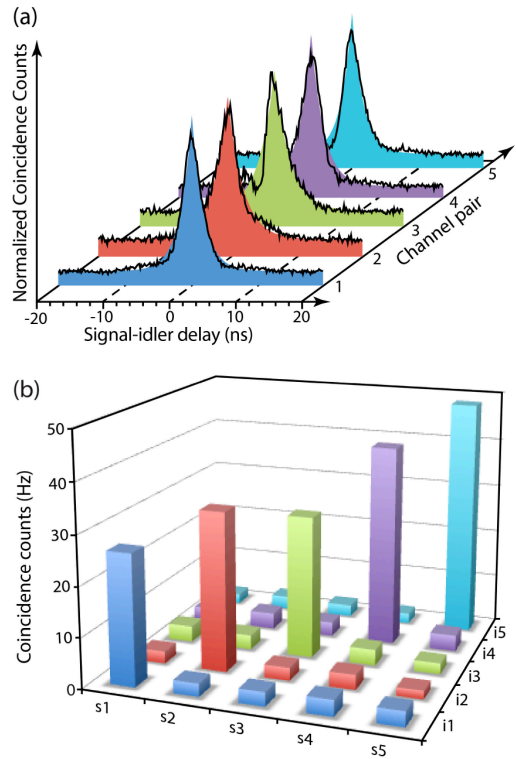

Fig. 2. (a) Measured coincidence peaks at five channel pairs centered around the pump wavelength. Clear coincidence peaks with a measured 110 MHz bandwidth (2.9 ns) are visible in all channel pairs. The solid-shaded curves are the $g^{(2)}$ fits of the experimental data (black curves). (b) Coincidence count rates measured at all the signal/idler combinations. Significant coincidence counts (corresponding to a peak) are only visible between symmetric channels (those with equal indices $m,n$ for both signal, $s_m$, and idler, $i_n$). For these measurements the idler detector was triggered by the signal detector (operating in free running mode).

We also estimated the heralded efficiency $\eta_h$, i.e. the probability of detecting an idler photon upon detection of the signal one. This definition does not take into account the quality of the heralded photon, thus in general a more complete definition should be taken into account [33]. However, since we determined a high purity of the heralded state with an independent measurement (see Sec. 2.3), we can evaluate the heralded efficiency as the ratio between the coincidence counts and the signal counts, in the configuration in which the idler detector is triggered by the signal one [34]:

$$\eta_h = \frac{cc}{c_{signal}\eta_{det}}, \qquad (1)$$

where $cc$ is the coincidence count rate, $c_{signal}$ is the signal count rate, and $\eta_{det}$ is the detection efficiency of the idler detector (5%). From the measured value we obtain a heralding efficiency (including all the losses) $\eta_h = 10\%$, limited primarily by the losses of the system.

We note that the pump power of 30 mW was chosen in order to optimize the CAR value while maintaining a reasonable count rate. Indeed, as shown in Fig. 3, the CAR decreases for increasing pump power. This is expected since higher pump powers increase the generation of multiple photon pairs, in turn increasing the accidental counts.

Even though our CAR values are not the highest reported in integrated structures [20] (partly a consequence of the losses in our detection systems), they are nonetheless suitable for many QKD applications [35]. A possible approach for reducing the losses of our system by a factor 6 dB, thus significantly increasing the performance, consists of exploiting a 2-port (instead of the 4-port structure used here) single bus waveguide microring. This, however, comes at the expense of making the self-locked scheme significantly more challenging to achieve.

The signal/idler temporal correlation is characterized by the Glauber second-order (cross) correlation function $g_{si}^{(2)}$ [36], that for a cavity takes the form [16]:

$$g_{si}^{(2)} = 1 + \frac{2\pi\Delta\nu}{2R}\exp\left(-2\pi\Delta\nu|\tau|\right), \qquad (2)$$

where $\Delta\nu$ is the ring linewidth, $R$ is the pair generation rate, and $\tau$ is the signal-idler delay. The experimental data [black curves in Fig. 2(a)] are well fitted by the $g_{si}^{(2)}$ function (solid-shaded curves) resulting in a measured value of $\Delta\nu_{fit}$ = 110 MHz, consistent with the linewidth of the ring resonator, after considering the time jitter of the single photon detectors

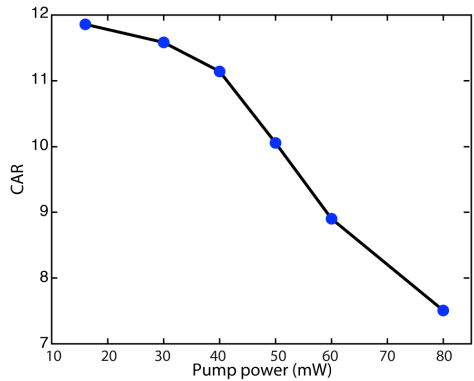

Fig. 3. CAR value as a function of the pump power (for the channel pair s5-i5).

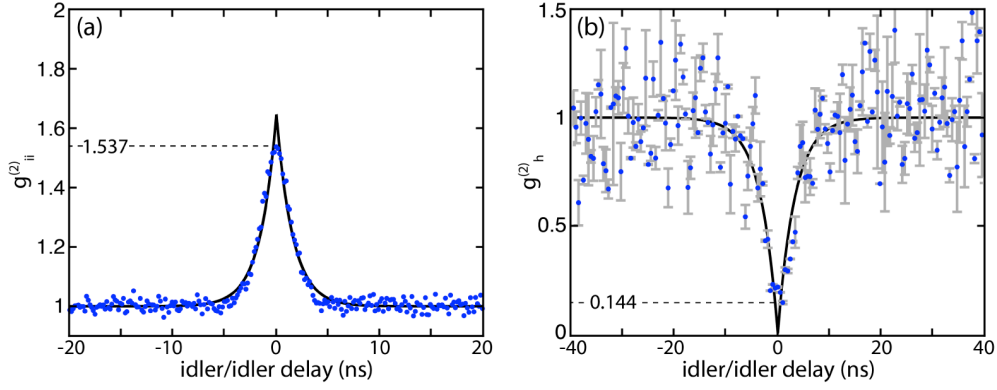

Fig. 4. (a) Idler-idler autocorrelation showing a bunching peak of 1.537 (corresponding to 1.86 effective modes). The black line represents the $g^{(2)}$ fit according to Eq. (1). (b) Heralded idler-idler autocorrelation. The antibunching dip of $0.144 \pm 0.008$ demonstrates the single photon character of the heralded source. The error bars are evaluated as the standard deviation on a 6-bin ensemble.

(810 ps, resulting in a corrected value $\Delta\nu_{corr}$ = 112 MHz). Considering that our source is nearly single-mode (see Sec. 2.3 for a detailed discussion) we can evaluate the coherence time as $\tau_{coh} = 1/(\pi\Delta\nu_{fit})$ = 2.89 ns [5].

The narrow photon pair bandwidth makes this device extremely appealing for applications such as quantum memories or quantum repeaters, where narrowband photon pairs are needed to match the atomic transition lines [13,37]. We note that the signal/idler bandwidths ($\Delta\nu$) of microcavity-based, CMOS-compatible sources reported to date, are inadequate for many applications, since they are orders of magnitude wider than what is required for quantum memories, due to the low microcavity Q-factors [6,8,18,19].

*2.3 Heralded single photon source*

In order to characterize our device as a heralded source of single photons, where the presence of a single photon is heralded by the detection of another photon, we measured both the single arm autocorrelation function $g_{ii}^{(2)}$ (on the idler channel H16) and the idler autocorrelation function heralded by the detection of a signal photon $g_h^{(2)}$.

The first quantity allows us to evaluate the purity of the source in terms of number of modes through the characterization of the statistics of a single beam (either the signal or the idler). For a spontaneous four-wave mixing (SFWM) source, as well as spontaneous parametric down conversion (SPDC), the pair statistics does not follow the pump statistics, which is typically a coherent field that exhibits Poissonian statistics, but it is instead thermal [38,39]. For an heralded single photon source with high purity the signal and idler spectra are uncorrelated within the linewidth of the microring, i.e. when considering signal and idler photons selected on the appropriate channels (symmetric with respect to the pump) no spectral information on the idler photon can be inferred from the measurement of the signal photon (or vice versa). Hence, the purity of the state is characterized by means of the joint two-photon spectrum, i.e. the measurement of the signal spectrum as a function of the idler wavelength. In our case the resolution of the available spectrometers was not enough to resolve spectral features within the microring linewidth. On the other hand such a narrow linewidth implies a long coherence time of the generated photons, thus allowing one to obtain the same information in the temporal domain through the measurement of the single arm autocorrelation function [22,40]. The purity of the state is in this case characterized by the number of the effective modes $N$: the lower $N$ the higher the purity. Since the statistics of the pair generation is thermal the number of modes is given by $N = 1/\left(g_{ii}^{(2)}(0)-1\right)$.

In the temporal domain the limitation is given by the time jitter of the detection system that is required to be smaller than the signal-idler coherence time (~810 ps and 2.9 ns in our setup, respectively). For a complete and detailed analysis of the effect of the temporal response of the detection system (with respect to the signal-idler coherence time) on the photon pair spectral factorability we refer the readers to [33]. For measuring the idler autocorrelation function we inserted a 50:50 fiber beam splitter on the H16 channel output and measured the coincidences between the two arms of the beam splitter. The results are shown in Fig. 4(a). We observed a bunching peak as high as 1.537, corresponding to $N = 1.86$ effective modes, in turn demonstrating the high purity of the proposed source.

As anticipated above, in order to demonstrate heralded single photon generation, we recorded the correlation between the output ports of the 50:50 beam splitter on the idler arm, conditioned by the measurement of a signal photon [41]

$$g_h^{(2)}(t_{i1}, t_{i2} | t_s) = \frac{P_{iis}(t_{i1}, t_{i2}, t_s)}{R^3 g_{si_1}^{(2)}(t_{i1} - t_s) g_{si2}^{(2)}(t_{i2} - t_s)}, \quad (3)$$

where $t_{i1(i2)}$ is the detection time of the idler photon at the first (second) beam splitter output port, $t_s$ is the detection time of the heralding signal photon, and $P_{iis}(t_{i1}, t_{i2}, t_s)$ is the triple coincidence rate.

For a perfect heralded single photon source we expect only one idler photon upon detection of a signal photon, and therefore we should observe a dip in the coincidence counts, indicating that the single idler photon is detected at only one of the beam splitter ports. In order to perform this measurement a single photon detector on the signal arm (channel H36) operating in free running mode (idQuantique id210) triggered the two detectors at the beam splitter outputs (idQuantique id210 and id200, set to 20 ns and 100 ns gate windows, respectively). From the time-tags of the three detectors we then extracted the heralded autocorrelation $g_h^{(2)}(\tau = t_{i1} - t_{i2})$. In order to account for the limited detection efficiency we introduced a heralding window $[-T_h, +T_h]$ (with $T_h = 0.81$ ns, thus ~4 times lower than the signal-idler coherence time) around the arrival time of the signal photon for accepting events on the triggered detectors [40]. The result, reported in Fig. 4(b), shows a clear dip of $g_h^{(2)}(0) = 0.144 \pm 0.008 = 0.5$, as expected for a source operating in the single-photon regime.

## 3. Discussion

Although we limited our investigation to five channel pairs, our device is capable of generating photon pairs on a comb over a bandwidth that exceeds the telecom C and L bands (1530 nm to 1620 nm), corresponding to more than 80 different signal/idler channel pairs. This conclusion stems from the recently reported results on above-threshold OPO obtained with the same device over a wide bandwidth (>300 nm, limited by our detector) [30,42]. Most importantly, in the unstable pump configuration, using a 50:50 fiber coupler at the drop port of the microring and two tunable filters, we directly observed a coincidence peak between signal and idler channels separated by over 100 nm, limited only by the available filters. In addition, while we focused on a particular pump wavelength, any resonant wavelength could have been chosen as the pump. For example, above threshold OPO operation was observed for different pump wavelengths, both in the TE and TM ring modes [30,42]. Note also that the microring resonances can be fine-tuned and shifted to any other desired frequency in the telecom band using temperature control (with a thermal frequency shift of –2 GHz/°C). This tunability could be relevant for matching the atomic transition lines for quantum memories.

The *multiplexed* nature of our device, enabled by the large phase matching bandwidth, has the advantage of maintaining the large data rates normally associated with ultrafast sources, while at the same time limiting the time-resolution required by available memories and detectors. It is worth noting that the multiplexed nature of our source is intrinsic in the physical process exploited (in the OPO process only the vacuum fluctuations at the frequencies satisfying the resonance condition can be effectively amplified), and it is not the result of a filtering procedure that, beside resulting in a strongly inefficient generation, is not suitable for most quantum applications [43].

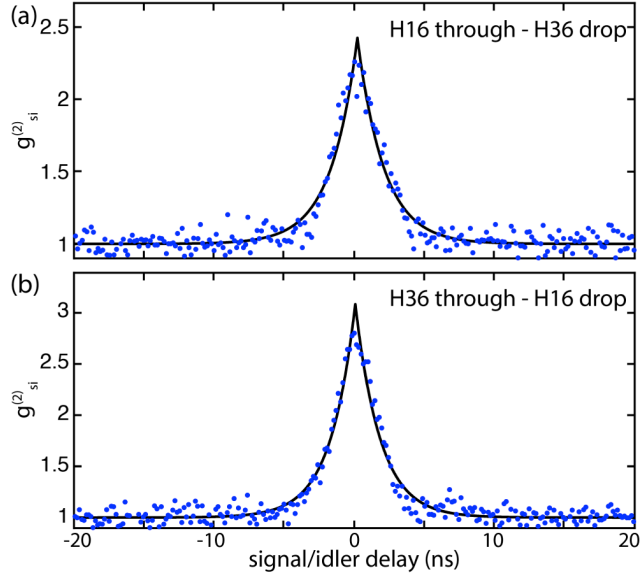

Fig. 5. Cross correlation functions obtained when extracting the photons from two different ring ports. (a) Cross correlation function measured when selecting the photons at 1564.3 nm (H16, i5) at the through port and at 1548.1 nm (H36, s5) at the drop port (blue dots), with the corresponding $g_{si}^{(2)}$ fit (black curve). (b) Same as in (a), but with the frequency filters inverted.

*3.1 Four-port microring*

In a four-port microring it is also possible to observe coincidences without spectrally separating the signal/idler photons, by exploiting the fact that the signal and idler photons have a 50% probability of exiting from different ports (drop and through). The photon pairs are generated inside the microring at the same time, but can exit the cavity independently, and have a 50% probability of exiting from the two distinct output ports (drop and through), while in the other cases they will exit both from the drop (25%) or both from the through (25%). It is therefore possible to directly separate signal and idler photons without relying on spectral filtering. The photons exiting from the drop port and transmitted by the 1x2 DWDM are sent to one detector, while the other detector collects the photons exiting from the through port (after removing the pump with a notch filter). In order to increase the pump rejection we added two frequency filters at the drop and through port, at frequencies symmetric with respect to the pump. The coincidence peak measured in this configuration is shown in Fig. 5(a). It is worth noting that the same peak appears even when the frequency filters are exchanged, further demonstrating that photons at conjugate frequencies exit from the two different ports of the ring [Fig. 5(b)].

*3.2 Shortening the cavity length*

Finally, we note that the pumping scheme used here can be further improved by replacing the EDFA with an amplifier having a shorter gain medium. In this way the external cavity length can be reduced until only a single external cavity line overlaps with a single ring resonance [31]. In this configuration stable CW operation can be achieved without the need for an additional Fabry-Perot filter. In order to demonstrate this, we replaced the EDFA with a commercial semiconductor optical amplifier (SOA, Thorlabs BOA1004S), which offered enough gain to sustain lasing, while only requiring a few centimeters of length. Using the fiber-coupled SOA as the gain medium, the external cavity was then shortened to a total length of 1.5 m, leading to an FSR of 135 MHz. As the new FSR was then close to the FWHM of the ring resonance, only one external cavity line oscillated within a single ring resonance at low pump powers, which resulted in stable CW pump oscillation. A coincidence peak was obtained also in this configuration (Fig. 6). These results clearly indicate that our pumping scheme has the potential to be fully monolithically integrated by incorporating the SOA and ring resonator on the same chip, an important achievement that would lay the foundation for on-chip quantum technology.

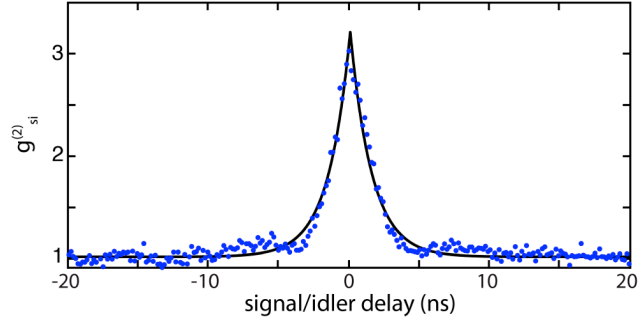

Fig. 6. Cross correlation function obtained when substituting the long EDFA amplifier with an SOA (blue dots) and the corresponding $g_{si}^{(2)}$ fit (black curve) for the $s_5$-$i_5$ channel pair.

## 4. Conclusions

We report a source of multiplexed photon pairs based on cavity-enhanced SFWM in a self-locked CW pumping scheme. We achieve a stable pump configuration, leading to extremely robust operation over several days without the need for any feedback or compensation for thermal shifts. The source is compatible with the ITU frequency channels for optical fiber communications networks and it achieves a very narrow photon pair bandwidth (~110 MHz) suitable for state-of-the-art quantum memories and repeaters. We demonstrate coincidence peaks over as many as five different signal/idler wavelength pairs, matching the standard ITU DWD communication channels, with the ability to generate pairs over the full L and C bands. We report nearly single-mode operation and demonstrate heralded single photons with a coincidence dip as low as 0.144. Our device is thus an ideal multiplexed source for many optical quantum applications, and is compatible with large-scale integration technology.

**Acknowledgments**

This work was supported by the Natural Sciences and Engineering Research Council of Canada (NSERC) and the Australian Research Council (ARC) Discovery Projects programs. C.R. acknowledges the support of a Vanier Canada Graduate Scholarship. L.C. acknowledges the support from the Government of Canada through the PDRF program and the "Fonds de recherce du Québec – Nature et technologies" (FRQNT) through the MELS fellowship program. We acknowledge the support from the People Programme (Marie Curie Actions) of the European Union's FP7 Programme: M.C. for the International Outgoing Fellowship (KOHERENT) GA n. 299522, M.F. for the International Outgoing Fellowship (ATOMIC) GA n. 329346, M.P. for the Career Integration Grant (THEIA) GA n. 630833, A.P. for the International Incoming Fellowship (CHRONOS) GA n. 327627. The authors thank Tudor W. Johnston, Yaron Silberberg, and William J. Munro for useful discussions as well as Ligthwave2020 Inc., Nick Bertone (Optoelectronics Components), Paolo Di Trapani, Amr Helmy, TeraXion, QGLex, and MPB Technologies Inc. for support. We are especially thankful to the referees for their rigorous and extremely useful review. Finally, we would like to thank John E. Sipe and Marco Liscidini for inspiring discussions.